\documentclass[runningheads]{llncs}
\usepackage{graphicx}
\usepackage{enumitem}
\usepackage{comment}

\usepackage{amsmath, amssymb,url}
\usepackage[utf8]{inputenc}
\usepackage[margin=1.1in]{geometry}
\title{U.S. Broadband Coverage Data Set: A Differentially Private Data Release}
\author{Mayana Pereira\inst{1}
\and 
Allen Kim\inst{1}
\and
Joshua Allen\inst{1}
\and
Kevin White\inst{1}
\and
Juan Lavista Ferres\inst{1}
\and
Rahul Dodhia\inst{1}}

\institute{Microsoft Corp.}
\date{}

\usepackage{graphicx}

\begin{document}
\maketitle

\begin{abstract}
Broadband connectivity is a key metric in today's economy. In an era of rapid expansion of the digital economy, it directly impacts GDP.  Furthermore, with the COVID-19 guidelines of social distancing, internet connectivity became necessary to everyday activities such as work, learning, and staying in touch with family and friends. This paper introduces a publicly available U.S. Broadband Coverage data set that reports broadband coverage percentages at a zip code-level. We also explain how we used differential privacy to guarantee that the privacy of individual households is preserved. Our data set also contains
error ranges estimates, providing information on the expected error introduced by differential privacy per zip code. We describe our error range calculation method and show that this additional data metric does not induce any privacy losses.
\end{abstract}

\section{Introduction}

Reaping the benefits of this digital world – pursuing new educational opportunities through distance learning, growing a small business by leveraging the cloud, and accessing better healthcare through telemedicine – is only possible for those with a broadband connection. As more people are staying home due to the COVID-19 pandemic, it becomes more evident the necessity of broadband connection for daily activities and better growth opportunities.

Based on the 2019 Broadband Deployment Report from the Federal Communications Commission (FCC) \cite{fcc}, broadband connection is not available to at least 21 million people in the United States, 16 million of whom live in this country's rural areas. Getting these numbers right is vitally important. This data is used by federal, state, and local agencies to decide where to target public funds dedicated to closing this broadband gap \footnote{https://www.fcc.gov/about-fcc/fcc-initiatives/bridging-digital-divide-all-americans}. That means millions of Americans already lacking broadband access have been made invisible, substantially decreasing the likelihood of additional broadband funding or much-needed broadband service. At the same time, making this data available in the clear, without any type of post-processing, may affect the privacy of individual households. In this contribution we solve this dilemma by presenting a data set that allows researchers and policy makers to develop solutions to improve broadband access while simultaneously preserving the privacy of individual households.

This paper describes the data privatization process used in the release of the broadband coverage data set\footnote{https://github.com/microsoft/USBroadbandUsagePercentages}. The data sets utilized are derived from anonymized data Microsoft collects as part of our ongoing work to improve our software and services' performance and security. We estimate broadband coverage by combining data from multiple Microsoft services. Every time a device receives an update or connects to a Microsoft service, we can estimate the throughput speed. We know the package's size sent to the computer, and we know the total time of the download. We also determine zip code level location data via reverse IP. Therefore, we can count the number of devices connected to the internet at broadband speed per each zip code based on the FCC's definition of broadband that is 25mbps per download \footnote{https://broadbandnow.com/report/fcc-broadband-definition/}. Using this method, we estimate that in 2020, $\approx 123$ million people in the United States are not using the internet at broadband speeds.

\textsc{Contributions:} . In this paper, we describe the U.S. Broadband Coverage Data set and the data privatization process to enable its safe sharing. Our two main contributions are:

\begin{itemize}[label = $\bullet$]

    \item We describe the privatization process of the Broadband Coverage Data Set. Our method utilizes differentially private mechanisms for data privatization. We describe implementation details and privacy loss specifications.
    
    \item We propose a method for error range estimation. Our empirical method estimates the error introduced by differential privacy for each zip code in our data set. The main advantage of our method is that it does not result in additional privacy losses.
    
\end{itemize}

\begin{figure}
\centering
\includegraphics[scale=0.5]{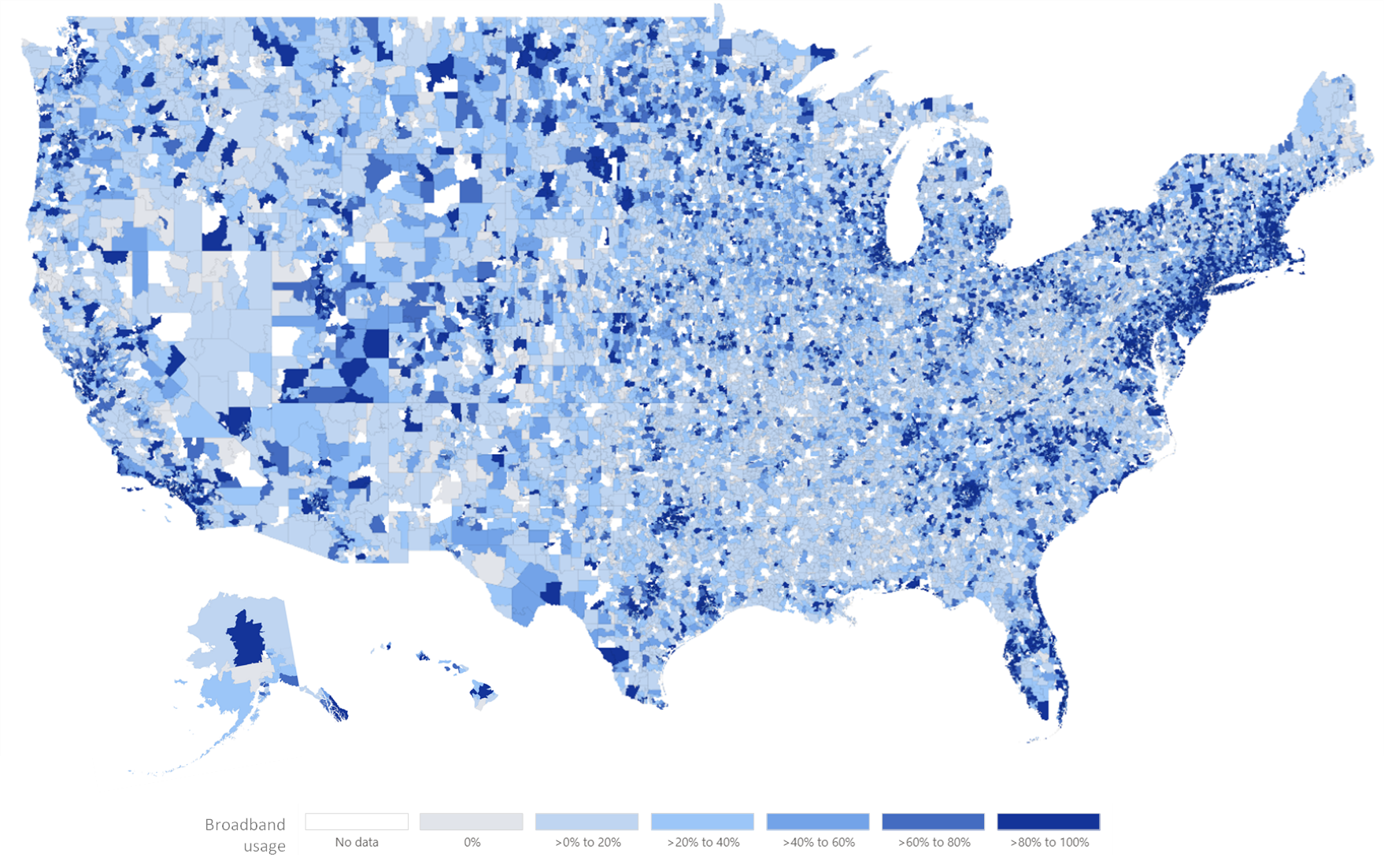}
\caption{Map of the United States by postal codes with indicators of broadband coverage.}
\label{fig:map}
\end{figure}

\section{Preliminaries}


Differential privacy is a rigorous privacy notion used to protect an individual’s data in a data set disclosure. We present in this section notation, definitions, and theorems that we will use to describe our privatization approach.  We refer the reader to \cite{book}, \cite{mcsherry} and \cite{calibrate} for more detailed explanations of these definitions and theorem proofs.

\begin{definition}{Differential Privacy} A randomized mechanism $\mathcal{M}:\mathcal{D}\rightarrow \mathcal{Y}$ with data base domain $\mathcal{D}$ and output set $\mathcal{Y}$ is $\epsilon$-differentially private if, for any output $Y \subseteq \mathcal{Y}$ and neighboring databases $D, D' \in \mathcal{D}$ (i.e., $D$ and $D'$ differ in at most one entry), we have
\end{definition}

$$
Pr[\mathcal{M}(D) \in Y] \leq e^{\epsilon}Pr[\mathcal{M}(D') \in Y]
$$

The privacy loss of the mechanism is defined by the parameter $\epsilon \geq$  0.

The definition of neighboring databases used in this paper is user-level privacy. User-level privacy defines neighboring to be the addition or deletion of a single user in the data and all possible records of that user. Informally, the definition above states that the addition or removal of a single individual in the database does not provoke significant changes in the probability of any differentially private output. Therefore, differential privacy limits the amount of information that the output reveals about any individual.

A function $f$ (also called query) from a data set $D \in \mathcal{D}$ to a result set $ Y \subseteq \mathcal{Y}$ can be made differentially private by injecting random noise to its output. The amount of noise depends on the sensitivity of the query.

\begin{definition}{$l_1$-sensitivity.} The $l_1$-sensitivty of a function $f: \mathcal{D} \rightarrow \mathbb{R}^n$ is:
\end{definition}
$$
\Delta f = max_{D, D'} \parallel f(D) - f(D')\parallel_{1}
$$

Where $D$ and $D'$ are neighboring databases.

We note that, for count queries, $\Delta f = 1$.

The Laplace distribution with 0 mean and scale $\lambda$, denoted by $\mathsf{Lap}(\lambda)$, has a probability density
function $\mathsf{Lap}(x|\lambda) = \frac{1}{2\lambda}e^{-\frac{x}{\lambda}}$.
It can be used to obtain an $\epsilon$-differentially private algorithm to answer numeric queries \cite{calibrate}.

\begin{definition}{Laplace Mechanism.} Let $f : \mathcal{D} \rightarrow \mathbb{R}^n$ be a numeric query. The Laplace mechanism is defined as:
$$
\mathcal{M}_L(x, f(\cdot), \epsilon) = f(x) + (\eta_1, \ldots,\eta_n) 
$$
 where $\eta_i$ are drawn from the Laplace distribution $\mathsf{Lap}(\frac{\Delta f}{\epsilon})$.
\end{definition} \label{laplace}

\begin{theorem} The Laplace Mechanism preserves $\epsilon$-differential privacy \cite{book}.
\end{theorem}

An important differential privacy property is its immunity to post-processing. The composition of a data-independent mapping $g$ with a $\epsilon$-differentially private algorithm $\mathcal{M}$, is also $\epsilon$-differentially private.

\begin{theorem}{Post-Processing \cite{book}} Let $\mathcal{M}$ be an $\epsilon$-differentially private mechanism and $g$ be
an arbitrary randomized mapping. Then, $g \circ \mathcal{M}$
is $\epsilon$-differentially private.
\end{theorem}\label{post}

\section{Differentially Private Broadband Coverage Estimates per Zip Code} 
The Broadband Coverage Estimates data set is derived from aggregated and anonymized information Microsoft collects as part of our ongoing work to improve software and service performance and security. 

Our metric for computing broadband coverage across the United States estimates the percentage of devices with internet connection speed over 25Mbps. We denote the set of zip codes in our data set by $\mathcal{Z}$. Our analysis uses zip codes as the smallest territorial unit. For every zip code $z$ in  $\mathcal{Z}$, the calculation encompasses the following variables:

\begin{itemize}[label=$\bullet$]

\item[] From Microsoft Services, we query the following Windows telemetry data:
    \subitem $L_{z}$: Counts of devices connecting to Microsoft Services with internet speed lower than 25Mbps in zip code $z$.
    \subitem $H_{z}$: Counts of devices connecting to Microsoft Services with internet speed greater or equal than 25Mbps in zip code $z$.

\item[] Additionally, we query Microsoft Services for the following data:
    \subitem $M_{z}$: Counts of devices utilizing Microsoft Services in zip code $z$.
    \subitem $O_{z}$: Counts of devices not utilizing Microsoft Services in zip code $z$.
\end{itemize}

Our data set contains 32,653 zip codes, covering most of the United States territory (see Figure \ref{fig:map}).

The Broadband Coverage Estimate ($\mathsf{BCE}$), is a metric that estimates the fraction of households of a zip code with access to internet with connection speeds over 25Mpbs. $\mathsf{BCE}$ is computed as follows:

\begin{equation}\label{bce}
   \mathsf{BCE}(z) = H_{z}\cdot(\frac{M_{z}}{M_{z}+O_{z}})^{-1} \cdot \frac{1}{\textnormal{HUD}_{z}} 
\end{equation}

Where HUD$_{z}$ is the number of households in zip code $z$, provided by the publicly available HUD data set \cite{hud}.
Given the zip code-level data set provides a granular view of broadband coverage, we ensure data privacy by utilizing differential privacy. For each of the counts queries from Microsoft Services Telemetry Data, we use the Laplace Mechanism, with an epsilon of $0.1$, to obtain differentially private counts. Our chosen privacy parameter is comparable to other industry differentially private data releases \cite{2020google} \cite{2020linkedin}. Because the Laplace mechanism can induce negative counts when the true counts are very small, we clamp all negative values to zero \cite{dp_18}. According to the post-processing immunity property (theorem \ref{post}), the privacy loss does not change by this step. We denote the differentially private counts of the variables used in our calculations by $L^{\textnormal{DP}}_{z}, H^{\textnormal{DP}}_{z}$, $M^{\textnormal{DP}}_{z}$  and $O^{\textnormal{DP}}_{z}$.

The published data set consists of a list of 32,653 zip codes. This list of zip codes was predefined based on average number of machines is the last years. For each zip code we provide the differentially private Broadband Coverage Estimate $\mathsf{BCE}^{\textnormal{DP}}$. $\mathsf{BCE}^{\textnormal{DP}}$ takes as input $L^{\textnormal{DP}}_{z}, H^{\textnormal{DP}}_{z}$, $M^{\textnormal{DP}}_{z}, O^{\textnormal{DP}}_{z}$ and is calculated according to equation \ref{bce}.

Our data set also includes the expected error, per zip code, induced by the privatization process. In section \ref{error} we describe the details of our error ranges estimation method.

\subsection{Privacy Loss}

The privacy loss computation is a straightforward application of the parallel and sequential composition properties of differential privacy mechanisms \cite{mcsherry}. The total privacy loss resulted from querying the internet speed telemetry is $\epsilon = 0.1$. Given that $L^{\textnormal{DP}}_{z}$ and $H^{\textnormal{DP}}_{z}$ are differentially private count queries applied to disjoint subsets of the data, from parallel composition we know that the privacy guarantee depends only on the worst of the guarantees of each analysis. The same happens when computing the privacy loss incurred from the Microsoft Services devices queries. The count queries are applied to disjoint subsets of data, resulting in an additional privacy loss of $\epsilon = 0.1$. From sequential composition we have that sequences of queries accumulate privacy costs additively. Finally, based on the post-processing immunity property described in theorem \ref{post}, the total privacy cost of the Broadband Coverage Estimates calculation is $\epsilon = 0.2$.

\section{Error Ranges Estimation}\label{error}
Differential privacy adds noise to protect privacy. A consequence of the additive noise is that differentially private counts from smaller subsets of data can be proportionally more affected by the noise, resulting in greater impact on utility. 

To ensure transparency into how zip codes with different population magnitudes are affected by differential privacy, we have included error range data. We propose an empirical methodology, via a simulation process, to calculate the expected error range caused by differential privacy. Although we can estimate differential privacy effects in query counts, the resulting error when combining several differentially private data sources is better estimated through a simulation process.

\begin{figure}
\centering
\includegraphics[scale=0.3]{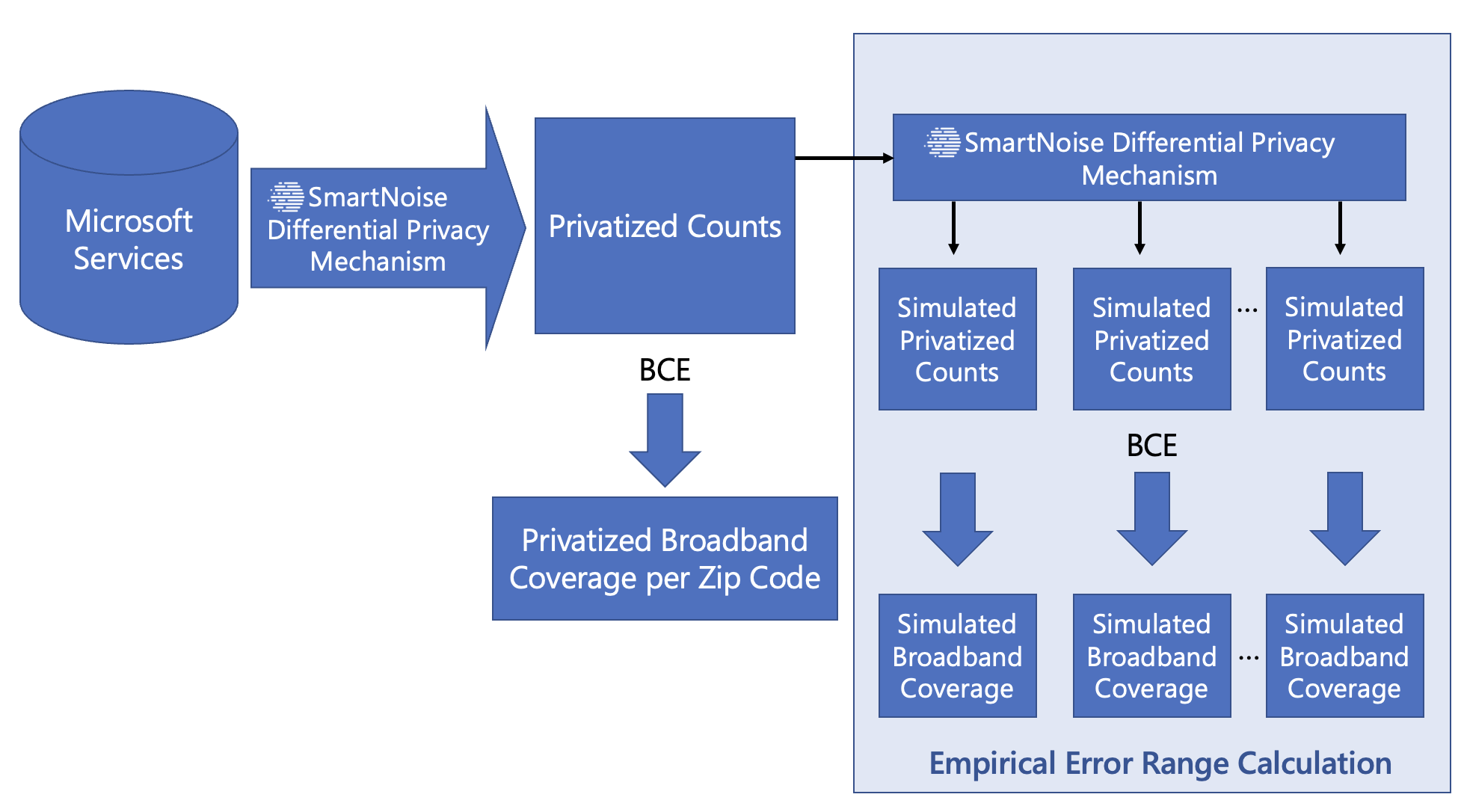}
\caption{Process for estimating the error introduced by differential privacy. The error ranges are estimated by generating several simulated private releases, and reporting the error introduced differential privacy.}
\label{fig:error}
\end{figure}

Our simulation process, illustrated in figure \ref{fig:error}, simulates
$k$ differentially private broadband coverage estimate $\mathsf{BCE}(z)$ for each zipcode $z$ in $\mathcal{Z}$. For each zipcode we estimate $95^{th}$ percentile error, mean absolute error and mean signed deviation.

All simulations occur in a post processing phase. For each zipcode $z$ in $\mathcal{Z}$, and for $i \in \{1,2, \ldots, k\}$, the error range estimation process occurs as follows:

\begin{itemize}
    \item $H^{\textnormal{DP}^i}_{z} \leftarrow H^{\textnormal{DP}}_{z} + \mathsf{Lap}(\frac{1}{\epsilon})$.
    \item $M^{\textnormal{DP}^i}_{z} \leftarrow M^{\textnormal{DP}}_{z} + \mathsf{Lap}(\frac{1}{\epsilon})$.
    \item $O^{\textnormal{DP}^i}_{z} \leftarrow O^{\textnormal{DP}}_{z} + \mathsf{Lap}(\frac{1}{\epsilon})$.
    \item $\mathsf{BCE}^{i}(z)$ takes as input  $H^{\textnormal{DP}^i}_{z}, M^{\textnormal{DP}^i}_{z}, O^{\textnormal{DP}^i}_{z}$ and is computed according to equation \ref{bce}.
    \item Finally, we compute $d^{z}_i = \mathsf{BCE}^{\textnormal{DP}}(z) - \mathsf{BCE}^{i}(z)$.
    
\end{itemize}

Each $\mathsf{Lap}(\frac{1}{\epsilon})$ is an independent sample from the Laplace distribution. For each zipcode, the simulation process will produce the vector $d^{z} = [d^{z}_1, d^{z}_2, \ldots, d^{z}_k]$, which consists of simulated errors caused by utilizing differentially private counts in the $\mathsf{BCE}$ calculation.

Finally, for each zipcode $z$ we estimate from $d^{z}$ the following error metrics: mean absolute error and $95^{th}$ percentile error. The non-private broadband coverage estimate will be, on average, within the mean absolute error (MAE) error range. Additionally, the $95^{th}$ percentile error gives a range where, for 95\% of the time, the non-private broadband coverage estimate will be within. We also provide the mean signed deviation (MSD). The mean signed deviation offers an estimate of bias introduced by the process. 

\begin{figure*}
\centering
\includegraphics[width=5.5in]{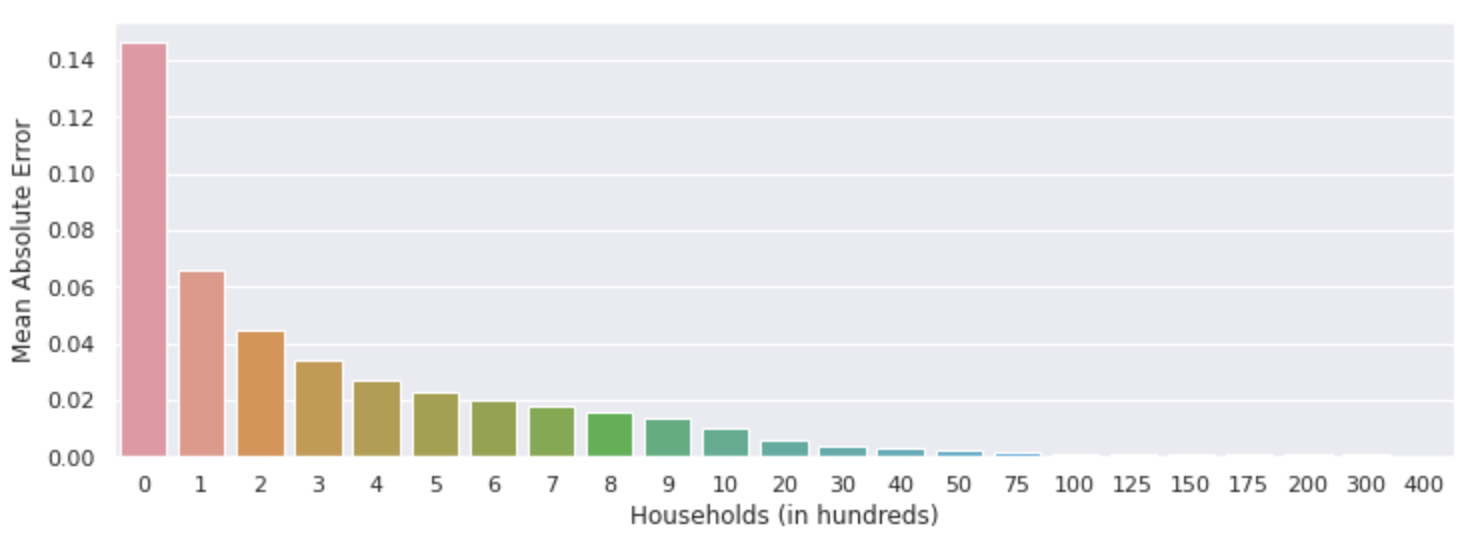}
\includegraphics[width=5.5in]{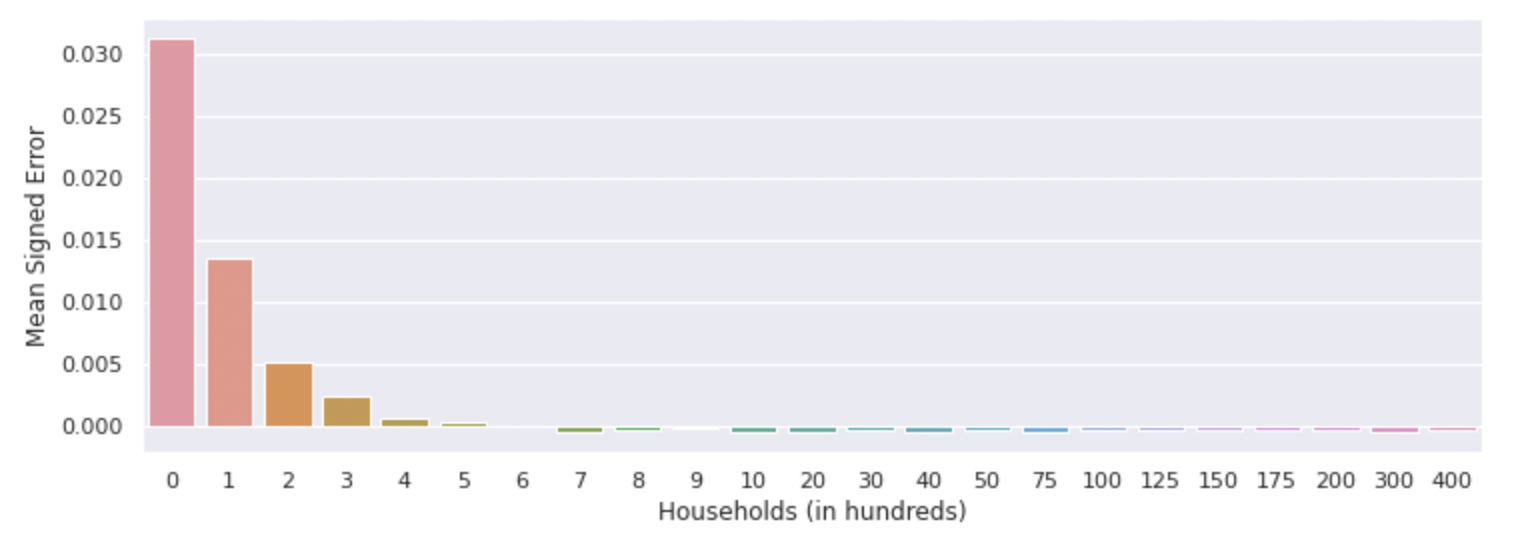}
\includegraphics[width=5.5in]{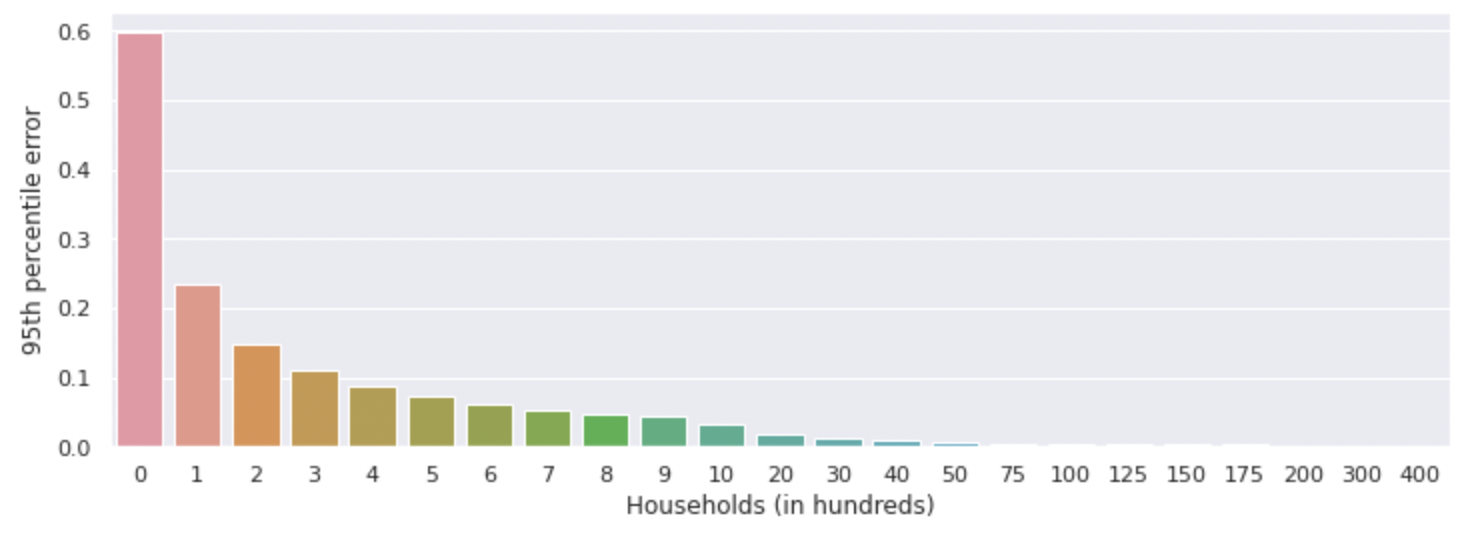}

\caption{Error metrics for several population ranges: Expected mean absolute error (MAE), mean signed error (MSD) and 95$^{th}$ percentile of the error caused by the privatization mechanism for different population thresholds.}
\label{fig_}
\end{figure*}

We display on Figure \ref{fig_} the error metrics for different population ranges (in number of households). Notice that the expected error decreases as the population size increases. Informally, we can say that the larger the population is, the less noise is needed to mask the presence or absence of an individual. Analogously, we can say that the smaller the population is, the more noise is necessary to mask the presence or absence of an individual.

The error range calculation is a post processing operation on the privatized data, i.e., all calculations use only the private counts $L^{\textnormal{DP}}_{z}, H^{\textnormal{DP}}_{z}$, $M^{\textnormal{DP}}_{z}$  and $O^{\textnormal{DP}}_{z}$ as input. From the post-processing immunity property, this process results in zero the privacy loss.
 
It is worth noting that the broadband coverage estimation has other sources of error, such as sampling error and bias. The error estimates provided in this data release do not account for such errors.  

\section{Implementation Details}
All differential privacy processing was done with the OpenDP SmartNoise library.  The SmartNoise library includes a comprehensive set of differential privacy mechanisms, algorithms, and validator.  The library is open source, and is maintained and vetted by OpenDP.

\section{Conclusion}
In this paper, we describe all the steps executed to ensure that the Broadband Coverage Estimates data set provides high utility data points while preserving privacy of Microsoft Services data. This data provides estimates of Broadband Coverage across the United States, at a zip code level. 

Additionally, we propose a simple method for estimating error via a simulation process. These supplementary data points help explain the impacts of differential privacy in different subgroup sizes. The main advantage of the proposed method for error generation is it does not cause any additional privacy losses.

\bibliographystyle{plain}
\bibliography{references}
\nocite{census, fcc,acc, tutorial}
\end{document}